%% file: root.tex
\documentclass[letterpaper]{article} % DO NOT CHANGE THIS
\usepackage[]{aaai25}  % DO NOT CHANGE THIS
\usepackage{times}  % DO NOT CHANGE THIS
\usepackage{helvet}  % DO NOT CHANGE THIS
\usepackage{courier}  % DO NOT CHANGE THIS
\usepackage[hyphens]{url}  % DO NOT CHANGE THIS
\usepackage{graphicx} % DO NOT CHANGE THIS
\def\UrlFont{\rm}  % DO NOT CHANGE THIS
\usepackage{natbib}  % DO NOT CHANGE THIS AND DO NOT ADD ANY OPTIONS TO IT
\usepackage{caption} % DO NOT CHANGE THIS AND DO NOT ADD ANY OPTIONS TO IT
\usepackage{algorithm}
\usepackage{algorithmic}
\usepackage{lipsum}
\usepackage{newfloat}
\usepackage{listings}
\DeclareCaptionStyle{ruled}{labelfont=normalfont,labelsep=colon,strut=off} % DO NOT CHANGE THIS
\floatstyle{ruled}
\newfloat{listing}{tb}{lst}{}
\floatname{listing}{Listing}
\usepackage{booktabs}
\usepackage{color,soul}%need to remove once done
\title{LLMs in the Classroom: Outcomes and Perceptions of Questions Written with the Aid of AI}
\author{
    Gavin Witsken, Igor Crk, Eren Gultepe\\
}
\affiliations{
    Dept. of Computer Science, Southern Illinois University Edwardsville, Edwardsville, USA\\
    gwitske@siue.edu, icrk@siue.edu, egultep@siue.edu
    
}

\usepackage{bibentry}
\begin{document}

\maketitle

% Uncomment the following to link to your code, datasets, an extended version or similar.
%
%\begin{links}
%     \link{Code}{https://aaai.org/example/code}
%     \link{Datasets}{https://aaai.org/example/datasets}
%     \link{Extended version}{https://aaai.org/example/extended-version}
%\end{links}
\begin{abstract}

This study evaluates the usage of OpenAI’s ChatGPT Large Language Model (LLM) as a tool for constructing multiple choice questions for assessing student academic performance through quizzes and exams.  
We randomly deploy questions constructed with and without use of the LLM tool and gauge the ability of the students to correctly answer, as well as their ability to correctly perceive the difference between human-authored and LLM-authored questions.  In determining whether the questions written with the aid of ChatGPT were consistent with the instructor's questions and source text, we computed representative vectors of both the human and ChatGPT questions using SBERT and compared cosine similarity to the course textbook. A non-significant Mann-Whitney U test (\textit{z} = 1.018, \textit{p} = .309) suggests that students were unable to perceive whether questions were written with or without the aid of ChatGPT. However, student scores on LLM-authored questions were almost 9\% lower (\textit{z} = 2.702, \textit{p} $<$ .01). This result may indicate that either the AI questions were more difficult or that the students were more familiar with the instructor’s style of questions. Overall, the study suggests that while there is potential for using LLM tools to aid in the construction of assessments, care must be taken to ensure that the questions are fair, well-composed, and relevant to the course material.

\end{abstract}

\input{intro}

\input{related}

\input{methodology}
\input{results}
\input{discussion}
\input{conclusion}

%\bibliography{aaai25}

\end{document}

%% file: intro.tex
\section{Introduction}

The workload of faculty at institutions of higher education in the United States is multifaceted, including requirements for teaching, research, service, and administrative duties. The workload varies depending on institution type, faculty rank, and discipline, but is generally considered demanding and includes responsibilities often extending beyond traditional working hours. 
%Faculty prepare and deliver courses, interact with students, grade assignments, conduct research, prepare work for publication, write grant applications, and perform peer reviews of unpublished work on behalf of journals and conferences. Additionally, faculty contribute to the administration of their institutions through various forms of service work. With these demands in place, the development of course materials for an upcoming term seldom entails fully designing a course and developing all materials from the ground up. In part, preparation for an upcoming term typically involves reviewing and revising assessment items to be used during that term. 
Writing assessment items can be especially time-consuming, since even with closely guarded materials, new questions must be constructed each term to ensure that students are being assessed on what they've learned this term, rather than what they've gleaned from those familiar with past iterations of a course. 
%Further, to make efficient the grading of assessments in heavily enrolled courses, faculty often rely on multiple choice questions (MCQs).

Multiple-choice questions (MCQs) can be a time-saving and effective tool for quick turnaround and prompt feedback to large groups of students. Well-designed MCQs can be used to assess not only recall of facts, but also understanding, application, and analysis. Grading MCQs is objective, reduces the potential for bias, and ensures consistency in evaluation. However, creating well-designed MCQs is a highly specialized task, requiring expertise not only in the knowledge or skill being assessed, but also in the methods used to derive an answer as well as plausible failures of those methods when they are not learned well. 

Haladyna provides an empirically validated set of guidelines for writing MCQs~\cite{haladyna2002review}, emphasizing clarity, relevance, and fairness in assessment. In summary: an MCQ consists of a stem (the part that poses the problem) and a set of possible answers; the stem should be clear and focused, avoiding unnecessary complexity and negative phrasing, while the choices should include distractors (incorrect options) that are plausible and which reflect common misconceptions that meaningfully challenge students; the choices should also ensure that the correct answer is unambiguous and that the options are similar in length and complexity to prevent clues. MCQs which follow Haladyna's guidelines may be considered well-designed.

This study is an investigation of using well-designed MCQs in a classroom, as written by the human instructor alone and through the use of a method proposed herein, for generating questions using AI with an expert human in the loop, validating AI output. We do not evaluate the efficiency of using AI to generate questions, or the quality of the generated output of LLMs, but instead focus on the assessment outcomes of validated, well-designed MCQs used in an actual classroom. We also evaluate student perceptions of the MCQs used in the course, asking them to predict each MCQ's authorship.

%% file: related.tex
\section{Related Work}

In recent years, standardized exams comprised of MCQs (LSAT, SAT, GRE, MedQA, MMLU, etc.), have been used to advertise the effectiveness of the current generation of AI tools~\cite{achiam2023gpt, gpt4o}. These datasets have become de facto standards for comparative evaluations of LLM performance. However, the history of the use of artificial intelligence in education, including the construction of well-formed MCQs, predates the recent ubiquity of LLMs.

Foundational approaches to the automatic generation of MCQs include work by Mitkov, et al.,~\cite{mitkov2006computer}, who construct an NLP method for generating MCQs while maintaining syntactic and semantic correctness, and relevance to the source text. G\"{u}tl, et al., also demonstrated the potential of AI in generating MCQs from textual content, specifically emphasizing the ability of AI to rapidly produce large question banks for large-scale educational settings~\cite{gutl2011enhanced}. 

More recently, the utility of LLMs for generating a large number of seemingly good quality MCQs has been remarked upon~\cite{khilnani2023potential}, as has the necessity for a human expert ascertaining their validity and reliability. This work is furthered by Rivera-Rosas, et al., where the authors conduct a study similar to our own, constructing 55 questions using ChatGPT3.5, of which 50 were used in an examination, with the remaining five eliminated due to redundancy or due to assessing material that is outside of the scope of the course~\cite{rivera2024exploring}. However, no other validation method is suggested and students were surveyed for their impressions about the language qualities of the examination (conciseness, clarity, complexity). Rivera-Rosas, et al., likewise stress that professors remain responsible for verifying the validity of the questions. However, whereas Rivera-Rosas, et al., evaluate an entirely LLM-authored assessment, our work provides a comparison of outcomes between faculty-validated AI-generated questions and their matched human-generated counterparts. We also suggest a validation methodology for LLM-authored MCQs. 

Cheung, et al., used human experts to assess the quality of 100 MCQs (50 authored by professors, 50 by ChatGPT)~\cite{10.1371/journal.pone.0290691}, with the assessors finding no statistically significant differences between the two sets of questions across five assessment domains. Their results suggest that a large number of good quality MCQs can be generated far more quickly with an LLM than by a human alone. Similarly, Olney demonstrated that there were no significant differences between MCQs written by humans or LLMs, when rated by medical experts for an anatomy and physiology textbook~\cite{olney2023generating}. The use of external assessors is undoubtedly valuable for standardized exams. Our own study asks the course instructor to validate questions solely for use in their own course.

%~\cite{10.1371/journal.pone.0305354}
%~\cite{eloundou2023gptsgptsearlylook}
%~\cite{tamkin2021understandingcapabilitieslimitationssocietal}

%% file: methodology.tex
\section{Methodology}

\begin{figure}[t]
\centering
 \includegraphics[width=.9\columnwidth]{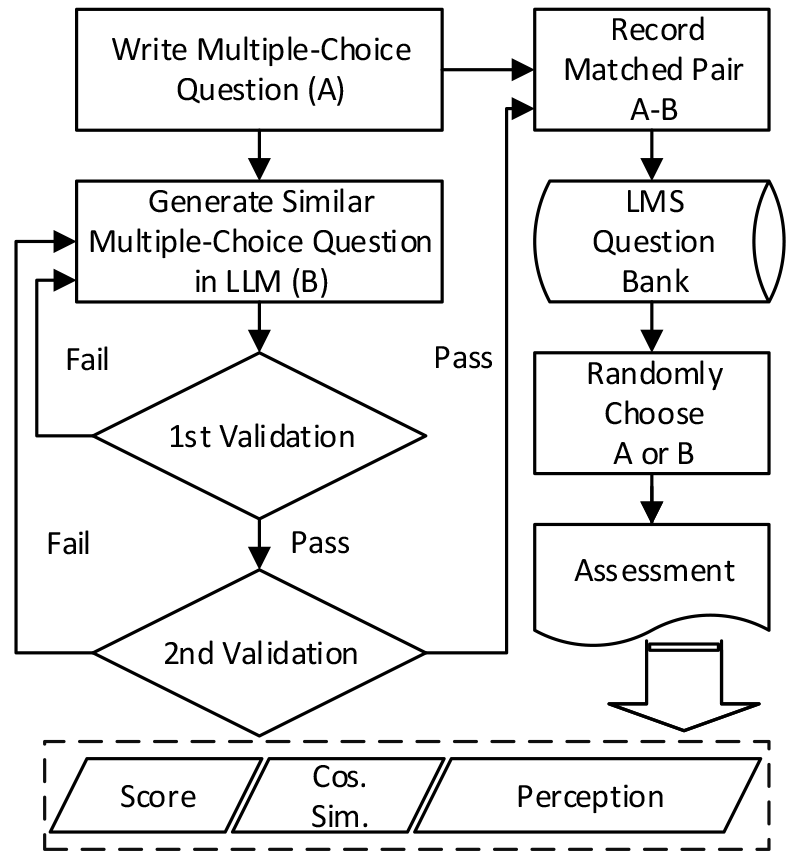} % Reduce the figure size so that it is slightly narrower than the column.
	\caption{Steps from question generation to data collection. We attempt to match each instructor-authored multiple-choice question (A) to a LLM-generated question (B). If (B) passes both human validation steps, it's added to the question bank along with the paired instructor-authored question. Each student's assessment includes either (A) or (B), chosen at random. Data collected includes the student's score, perception (whether question appears to be human- or LLM-authored) and the cosine similarity of the question and the course's textbook.}
	\label{flowchart}
\end{figure}

We consider three key questions in the design of the experiments: 
\begin{enumerate}
\item For assessing student learning, does the use of an LLM as a tool for constructing MCQs affect student scores? 
\item Can students distinguish between MCQs written with the aid of an LLM from those that were not? 
\item Can a difference (or lack thereof) between the two types of questions be quantified and would it be reflected in either student outcomes or perceptions of the questions? 
\end{enumerate}

To address these guiding questions, we designed and deployed an experiment in a third year operating systems course at Southern Illinois University Edwardsville (SIUE) to examine student performance on assessment questions constructed with the use of an LLM and ask the students to rate the likelihood that each question was written by a human alone.

The 25 students enrolled in the 8-week synchronous online course were traditional students in their third year of undergraduate study at SIUE. As is typical of operating systems courses, students were expected to complete several programming assignments and a number of quizzes and examinations. For the purpose of this study, we considered their performance on 32 questions deployed across four quizzes and two exams, as these are performed in a relatively controlled environment; i.e. student performance on programming assignments is not considered here. The range of possible scores for each question is [0, 100]. Of the 32 traditionally constructed questions, 24 had counterparts that were constructed with the aid of an LLM and matched to the topic the traditionally constructed question was written to assess. Questions with matched counterparts were randomly deployed, such that each student was exposed to some LLM-aided questions over the course of the term. Student responses are recorded as correct, partially correct, or incorrect. Course contents and assessments were deployed using the Moodle LMS~\cite{moodle}.

\begin{table*}
\fontsize{9pt}{9pt}\selectfont
\centering
\renewcommand{\arraystretch}{1.5}
\begin{tabular}{p{2.5cm} p{14.25cm}}\toprule
\textbf{Question Generation Steps}  & \textbf{Sample Input/Output Text} \\
\hline
\textbf{1.} Instructions to ChatGPT & \textbf{Input}: "This GPT specializes in generating quiz questions for students in an undergraduate Operating Systems course. It should provide well-structured, clear, and appropriately challenging questions covering various topics such as processes, threads, CPU scheduling, memory management, file systems, and security. It should be able to create multiple-choice, true/false, and short-answer questions, ensuring a range of difficulty levels. The goal is to help students practice and test their understanding of Operating Systems concepts." \\
\hline
\textbf{2.} Prompt to ChatGPT & \textbf{Input}: Create a multiple choice question which requires students to identify how many disk operations are required to fetch the inode of a file at some path. Please clearly state any assumptions or clarifications that would be necessary to identify this. Note that reading a block and fetching an inode are considered 2 distinct disk operations.\\
\hline
\textbf{3A.} Generated MCQ which failed validation & \textbf{Output}: Consider a file system that uses an inode-based structure similar to Unix. Assume the following: The file system's directory entries contain the name of the file and a pointer to the inode. Each directory block can hold a maximum of 10 entries. The file system has a multi-level directory structure where each level is stored in a separate block. Reading a block and fetching an inode are considered two distinct disk operations. Given a file path /dir1/dir2/file.txt, how many disk operations are required to fetch the inode of file.txt? A) 2 B) 3 C) 4 D) 5.\\
\hline
\textbf{3B.} Generated MCQ which passed validation & \textbf{Output}: Assume you have a file system with a hierarchical directory structure, where inodes are stored separately from data blocks. Each directory entry contains only the name of the file or subdirectory and a pointer to its corresponding inode. Consider a file located at the path /dir1/dir2/file.txt. The root directory inode is always in memory, but all other inodes and data blocks must be read from disk. How many disk operations are required to fetch the inode of the file file.txt at the given path? Clarifications: Each disk operation is defined as either reading a block from disk or fetching an inode from disk. The block containing directory entries and the inode itself are separate entities. The root directory inode is already in memory and does not require a disk operation. Choices: A) 2 B) 4 C) 6 D) 8 \\
\bottomrule
\end{tabular}
\caption{A sample input and output procedure for generating LLM-based questions. 1) Overall instructions are provided to contextualize ChatGPT's generation of all MCQs. 2) A topic-specific prompt requesting a multiple choice question is then given. 3A) An example of a generated question that failed two-pass validation, due to having unstated but necessary assumptions. In the absence of these necessary assumptions, the MCQ stem is ambiguous, resulting in several correct answers, some of which were not even given as a choice. 3B) A generated question that passed both stages of validation, since the question is unambiguous and has the single correct answer: C. } 
\label{table:examples}
\end{table*}

\begin{figure*}[ht]
\centering
 \includegraphics[width=.80\textwidth]{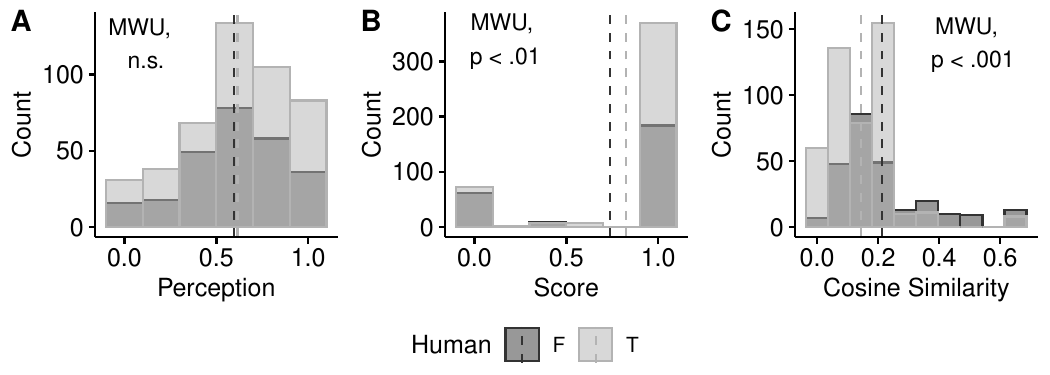} % Reduce the figure size so that it is slightly narrower than the column.
	\caption{Histograms showing the effect of human written and AI generated questions on student performance. A) A Mann-Whitney U (MWU) test (\textit{z} = 1.02, \textit{p} = .31) showed that there was no difference in student's ability to detect human (\textit{M} = 0.62, \textit{SD} = 0.28)  from AI (\textit{M} = 0.60, \textit{SD} = 0.27) questions. B) A MWU (\textit{z} = 2.70, \textit{p} $<$ .01) showed that students scored higher on human (\textit{M} = 0.83, \textit{SD} = 0.37) written questions rather than AI-generated ones (\textit{M} = 0.74, \textit{SD} = 0.43). C) A MWU (\textit{z} = -5.89, \textit{p} $<$ .001) showed that human written questions (\textit{M} = 0.14, \textit{SD} = 0.11) were significantly less similar to the course textbook than the AI-generated ones (\textit{M} = 0.21, \textit{SD} = 0.15).}
	\label{fig2}
\end{figure*}
All MCQs were presented alongside an ungraded and optional response which asked the student to rate the likelihood that the question was written entirely by a human; the response for this rating was captured using a 6-point scale with options being 0\%, 20\%, 40\%, 60\%, 80\%, and 100\%, purposefully omitting a 50\% option, so that each respondent must carefully consider whether their perception leans more  towards human or towards AI when they are unsure of the question's authorship. 

\subsection{Question Selection and Validation}

For each quiz or exam question that is written by the instructor, an alternative question was included to be randomly deployed instead of the instructor's own only if the LLM (ChatGPT) was capable of constructing a question that passed the instructor's two-pass validation. The flowchart in Figure~\ref{flowchart} illustrates our methodology. 

In the first pass, MCQs were considered acceptable if they gave clear and unambiguous directions, if they did not give unintended clues, and if they appeared to test the knowledge or skill in a way that matched the intent of the instructor's original question. 

In the second pass, we worked out the solution to the candidate question with an eye on the provided choices, ensuring that the expected student effort was similar to what was expected by the instructor's original question, that the choices also presented responses that were incorrect but plausible in the presence of a flaw in reasoning as might be expected from an underprepared student, and that the correct choice was present and could clearly be reasoned from the MCQ's stem. In other words, we use the same analysis and criteria we apply to assessing the effectiveness of our MCQs to assess the effectiveness of questions written with the help of an LLM. Table~\ref{table:examples} shows one example of a question that was rejected in the second pass: while the question appears to be well formed at first glance, working out the solution reveals that critical assumptions are not stated, so several plausible answers exist, some of which are not given as choices. 

Since we were not measuring the instructor's effort or the differences in instructor effort that may present when using an LLM to construct assessments, and because we suspect that any prompt iteration will drive the output to be more indistinguishable from the instructor's own choice of words or phrases, we strictly used zero-shot prompting, strictly avoiding iteration. If the output did not pass validation, the entire conversation was abandoned and the same prompt given again in a new conversation; this was repeated until an acceptable (i.e. passing validation) question was constructed by the LLM. For many questions this procedure was repeated several times before an acceptable question was found. Once an acceptable question was found, it was deployed exactly as constructed by the LLM. 

It should be noted that of the 32 instructor written questions, 24 had counterparts written with the aid of an LLM. Of the eight missing ones, three are missing counterparts due to simple time constraints and five did not pass validation after many attempts to generate a matching question of similar quality and expected effort to the instructor's own question. 

\subsection{Other Constraints}

We minimize the risk to student educational outcomes and preserve course quality and rigor by only using those questions that are validated by the instructor. Furthermore, as the study was conducted in a real course with real stakes, each student was randomly given either the instructor's own question or its LLM-aided, validated counterpart. The random assignment of questions is done for each student and for each question. Since the integrity of the course and its educational outcomes is a priority, this experimental design ensured that no unforeseen consequences were concentrated on some unlucky group of students. 

Students were not told whether their perceptions of each question's authorship was correct or not. No additional time was required of or granted to the students beyond what is normally given for quizzes and exams. Quizzes were completed online, prior to lecture, with 5 minutes allotted per question. Exams were completed during scheduled class meeting times and were allotted no more than the scheduled meeting time.

\subsection{Similarity of Questions to Course Material}
Intuition suggests that the instructor's assessments will resemble the course material in style as well as substance, and the LLM output may not. It would also be reasonable to expect that if the textbook is part of the LLM's training data, its questions may be a closer match to the text than the instructor's own creative constructions.

A fair assessment in a classroom setting considers what was emphasized, so in preparation and prior to lecture, students were expected to have completed assigned readings from at least one of the course's required textbook (Andrew Tanenbaum's \textit{Modern Operating Systems, 4th Edition} was used in this course). The quizzes in particular were constructed based on choice end-of-chapter questions, when appropriate, and significantly rewritten and expanded into MCQs, in which case the wording of the instructor's MCQs do not necessarily appear in the text. Other questions were devised to reflect lecture material not necessarily based on the assigned readings. The LLM-authored questions arise out of the underlying model's training data, which is likely to include the contents of the required textbook ~\cite{sitnflash_making_2023}. In contrast, exam questions are reflective of the assigned readings as well as the contents of programming projects. 

We anticipated that any differences between LLM-authored MCQs and lecture contents might cause a difference in either student performance or their perception of the authorship of the questions. To gain some insight into any performance or perception differences which may arise, we attempted to quantify the relationships between the course text and MCQs of both types by calculating cosine similarities between them using SBERT embeddings ~\cite{reimers-2019-sentence-bert}.

\subsection{Protocol Approval}

The experimental protocol was approved by the institutional review board of SIUE (IRB protocol \#2591). Students were informed of the study, protocol, and informed that their participation was voluntary and that, as such, non-participation or withdrawal from the study carried no penalty. 

\subsection{Statistical Analysis}
In our analysis, Perception, Score, and Cosine Similarity were considered to be  predictors of question authorship. The histograms in Figure~\ref{fig2} show the distribution of each variable, grouped by the ground truth of MCQ authorship (denoted in the figure as Human, with \textit{True} indicating that the question was authored by a human and \textit{False} indicating that it was authored by LLM, i.e. ChatGPT). Due to the nonlinearity of the data, we used Mann-Whitney U tests to determine whether any statistically significant differences exist within each of the three variables when authorship is considered.

To determine the presence of interactions among the statistically significant variables, binary classification on the questions was performed in which the Human variable was considered as the ground truth. The baseline classification method was a logistic regression (LR) with the interaction terms. To get a better understanding of the possible interactions among the variables, a Conditional Inference Tree (CIT) was constructed~\cite{hothorn2015ctree}. CITs have been shown to successfully parse high-order interactions among two or more variables~\cite{levshina2021conditional}, since effects of the predictor variables are considered simultaneously. This can be attributed to its use of \textit{p}-values when performing splits on the predictor variables. CIT was trained using grid search. Both classifications were performed using five-fold cross-validation (CV) and the area under the receiver operating characteristic curve (AUC) was used as the performance metric.   

\begin{figure}[!h]
	\centering
	\includegraphics[width=0.85\columnwidth]{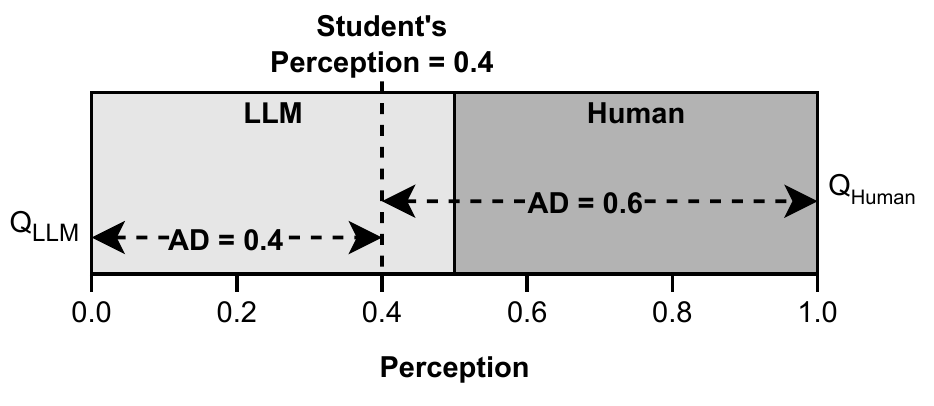} 
	\caption{If a student's Perception value is 40$\%$ and the true question authorship is human, then the AD is 0.6 and if the true authorship is LLM, then the AD is 0.4.}
	\label{fig_per}
\end{figure}

Also, to highlight any patterns among students regarding their ability to correctly perceive the an MCQ's authorship, hierarchical clustering using Euclidean distance and Ward linkage, was performed on the absolute difference (AD) between a student's perception and the ground truth. For example, if a student selected 40\% likelihood that the MCQ was authored by a human and the MCQ was indeed authored by a human, the AD would be 60\% and if the MCQ was authored by the LLM, the AD would be 40\% (see Figure~\ref{fig_per} for a visual example).  The quality of the clusters was determined by the silhouette values, where between 0.3 and 0.5 demonstrates reliable clustering, $\geq$ 0.50 demonstrates good clustering, and $\geq$ 0.70 indicates strong structure; the best value possible is 1 and the worst is -1. All statistical analyses were independently verified using R, Python, and IBM's SPSS.

\begin{table}[t]
\centering
\begin{tabular}{ l c c c }\toprule
\textbf{Metric}  & \textbf{N} & \textbf{Mean} & \textbf{St.Dev.} \\
\hline
Score & 714 & .793 & .394 \\
Score (human) & 459 &.827 & .367\\
Score (LLM) & 255 & .739 & .429 \\
\hline
Perception & 714 & .625 &.285 \\
Perception (human) & 459 & .615 & .285\\
Perception (LLM) & 255 & .598 & .271\\
\hline
Cosine Similarity & 714 & .168 & .127\\
Cosine Similarity (human) & 459 & .144 & .105\\
Cosine Similarity (LLM) & 255 & .212 & .150\\
\bottomrule
\end{tabular}
\caption{Counts, means, and standard deviations of collected data (assessment scores, perception scores, and cosine similarity of questions to the course textbook).}
\label{table:descriptives}
\end{table}

%% file: results.tex
\section{Results}

Responses were collected from up to 32 MCQs for each of the 25 enrolled students. Two students' responses were ultimately omitted as outliers; one student had dropped the class early in the term and the other had rated their perception the same for all questions. Of the 32 questions written by the instructor to assess students, 24 had AI-generated counterparts, which were used to randomly substitute the instructor's matching question at random.  Out of a possible 736 data points, we collected 714, indicating that a few students had skipped some questions. 

\subsection{Predictors of Question Authorship}

Table~\ref{table:descriptives} details the basic descriptive statistics of the collected data. Score, Perception, and Cosine Similarity were considered as predictors of MCQ authorship in our experimental design. Score indicates the overall graded outcome of all 714 MCQs used. For example, the overall course average for the quiz and exam questions used here was 79\%.  When separating Score by authorship, it is apparent that the mean of scores for human-authored MCQs (\textit{M} = 0.83, \textit{SD} = 0.37) is almost 9\% higher than the mean of LLM-authored MCQ scores (\textit{M} = 0.74, \textit{SD} = 0.43) and shown to be significantly the different (Figure~\ref{fig2}B) with a Mann-Whitney U test (\textit{z} = 2.70, \textit{p} $<$ .01).

Perceptions in Table~\ref{table:descriptives} represent the students' certainty that a question was written in its entirety by a human (on a 6-point scale, with a 0\%-100\% range and 20\% intervals). Perception scores are also separately shown for human-authored and LLM-authored questions and illustrated in Figure~\ref{fig2}A. A Mann-Whitney U test (\textit{z} = 1.02, \textit{p} = .31) showed that there was no difference in student's ability to detect human (\textit{M} = 0.62, \textit{SD} = 0.29)  from AI (\textit{M} = 0.59, \textit{SD} = 0.27) questions. 

Lastly, in Table~\ref{table:descriptives}, the Cosine Similarity scores represent the similarity of each question's text to the textbook. Interestingly, LLM-authored (\textit{M} = 0.21, \textit{SD} = 0.15) questions appear to be more similar to the textbook than the human-authored ones (\textit{M} = 0.14, \textit{SD} = 0.11). A Mann-Whitney U test of the difference illustrated in the histogram in Figure~\ref{fig2}C was significant (\textit{z} = -5.89, \textit{p} $<$ .001).

\begin{figure*}[!ht]
\centering
 \includegraphics[width=.96\textwidth]{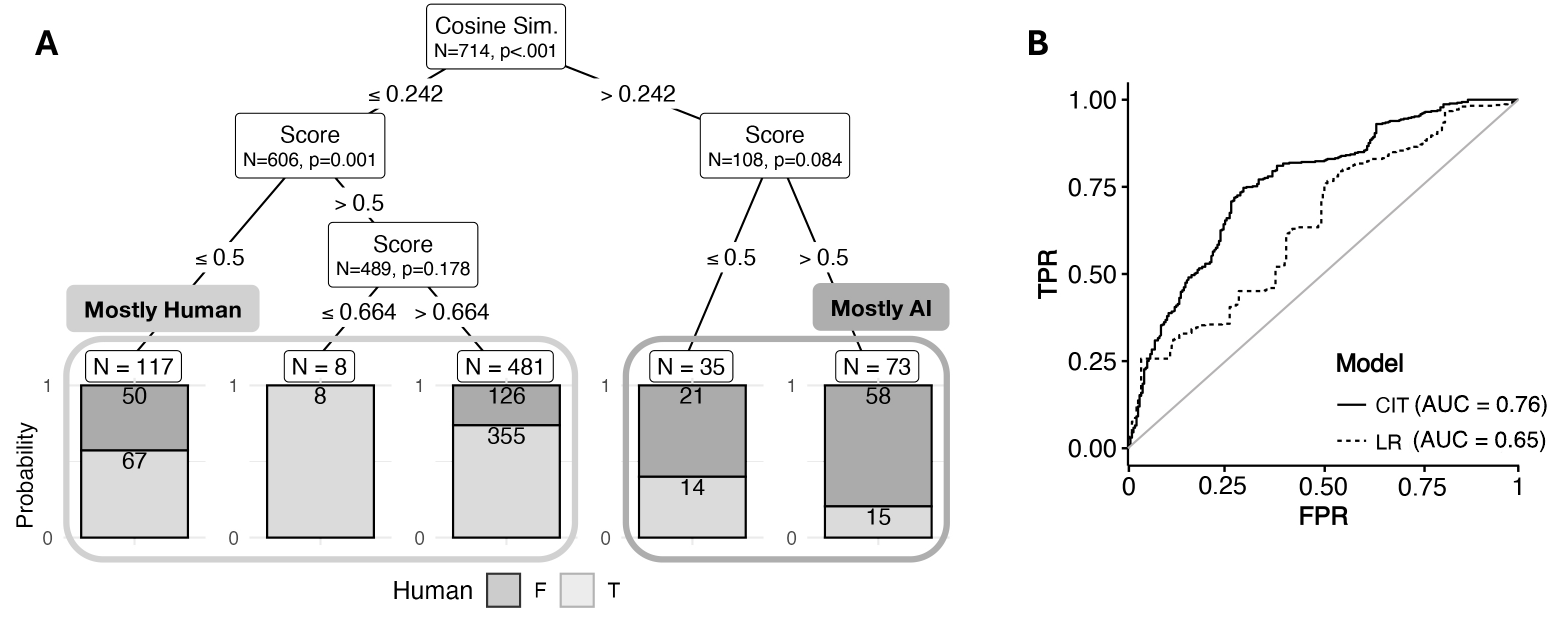} 
	\caption{A) The tree shows the interaction between cosine similarity and score in ascertaining whether a question is authored by a human or AI. Due to the negative correlation ($r_{s}$ = -.13, \textit{p} $<$ .001) between cosine similarity and human-authorship, we can expect that when cosine similarity is $\leq$ 0.242, more human-authored questions will be represented (light-gray portion of leaf nodes), and otherwise we expect to see more AI-authored questions (dark-gray portion of leaf nodes).  Based on this separation of MCQs, we can also compute that when cosine similarity is $\leq$ 0.242 students scored better on human written questions (mean error rate of 0.23) than AI generated questions (mean error rate of 0.303). B) CIT had a higher (AUC = 0.76) than LR (AUC = 0.65) for determining question authorship, which could be due to better modelling of the interaction of the Cosine Similarity and Scores.}
	\label{fig3}
\end{figure*}

Using the significant predictors (Cosine Similarity and Score) as determined by the Mann-Whitney U tests, classification of the question authorship (Figure~\ref{fig3}) using LR (Figure~\ref{fig3}B, AUC = 0.65) showed that the were was statistically significant interaction between Cosine Similarity and Score (\textit{p} $<$ 0.05). With the CIT model (Figure ~\ref{fig3}A), the classification  performance increased (Figure~\ref{fig3}B, AUC = 0.76) and interaction between both predictors was observed at a higher granularity. Specifically, Figure \ref{fig3}A shows that the interaction between Score and Cosine Similarity depends on a statistically significant (\textit{p} $<$ .001) split occurring at a Cosine Similarity value of 0.242, below which students scored better on human-authored questions, reflecting the negative correlation between Cosine Similarity and Score ($r_{s}$ = -.13, \textit{p} $<$ .001).

\begin{figure}[!ht]
	\centering
	\includegraphics[width=0.92\columnwidth]{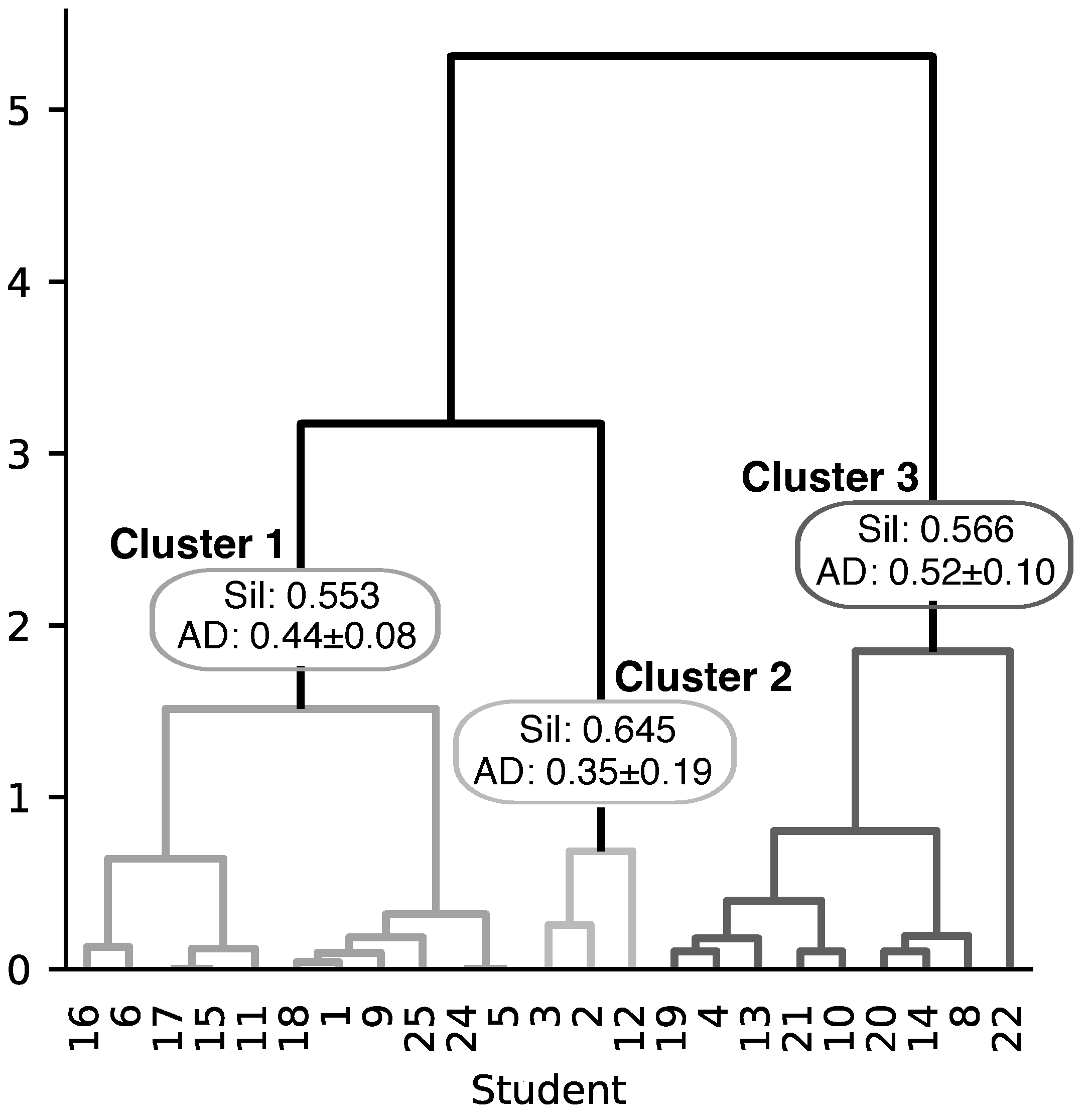} 
	\caption{Hierarchical clustering using each student's average absolute difference (AD) values strongly suggests that there are three  distinct groups of students (silhouette = 0.570). Students in cluster two had the best ability in differentiating authorship among MCQs.}
	\label{fig4}
\end{figure}

\subsection {Student Perceptions vs Question Authorship}

Hierarchical clustering (Figure~\ref{fig4}) demonstrated that there were three major groups of students (silhouette = 0.570). The three-cluster organization demonstrated that a group of students (cluster two: \textit{M} = 0.351, \textit{SE} = 0.190, sil. = 0.645) were very likely to identify the correct authorship of a question irrespective of it being written by a human or generated by an LLM. Cluster one had a mean AD that indicated the correct authorship was being attributed by the students (\textit{M} = 0.440, \textit{SE} = 0.075, sil. = 0.553), just not as strongly as cluster two. Students in cluster three (\textit{M} = 0.519, \textit{SE} = 0.102, sil. = 0.566) where more likely to incorrectly attribute the authorship, since the mean AD was higher than 0.5. The dendrogram in Figure~\ref{fig4} was cut at other heights to investigate other clusters structures, but the overall silhouette scores were lower than the chosen three-cluster solution. %The following were the silhouette values obtained using the other clustering structures: two clusters) sil. = 0.529; four clusters) sil. = 0.565; and 5 clusters) sil. = 0.567.

%% file: discussion.tex
\section{Discussion}

\subsection{Effect of Authorship on Student Performance}

Students performed better on human-authored MCQs than the LLM-authored ones. While the LLM-authored MCQs were validated and considered usable if they were relevant to the topic, the instructor's matching question was intended to assess the validation process and intentionally did not include any consideration of whether the phrasing, tone, apparent approach to a topic was a close match to the instructor's. It could be reasonably expected that further prompting or prompt refinement could bring the AI-generated text into alignment with the instructor's own personal phrasing, tone, etc., and diminish these differences. The practice of using only the LLM's first response was expected to emphasize the possible differences. 

\subsection{Perception and Similarity} 

The perception scores were examined for any obvious signs of outliers, finding two such students, one of whom always gave the same perception rating to each question and another who only partially completed the course. Based on the overall results, the students did not perceive a difference between human-authored and AI-authored questions. This was not entirely unexpected, given that AI-generated images, videos, and text are usually flawed in way, which allows people to discern that some artifact is indeed AI-generated. In ensuring, through validation, that the questions used in this study did not adversely affect measurement of student learning and were otherwise reasonable, we relied on the human in the loop to remove such obvious flaws. The basis on which the students made their authorship determination is unknown and worth exploring.

Cosine similarity calculations showed that the human-authored MCQs were significantly less similar to the course textbook than the AI-generated ones. While surprising at first, it is our opinion that it may be reasonable to expect that a textbook published in 2014, easily found online in both text-only and PDF formats, and often discussed in various online forums, would be part of a large language model's training data. It is also reasonable to expect that introductory textbooks are similar in their use of established jargon and phrasing, and share a focus on the fundamental concepts of a particular subject. In contrast, the instructor's questions were novel MCQs.

One plausible explanation for the significant interaction between Scores and Cosine Similarity is that while the LLM-authored MCQs are more similar to the language of the textbook, students may not have read the assigned text, relying instead on lecture notes and other course materials to complete the assessments. As this data was collected in an online synchronous course, where quizzes and exams were also given online, it may be expected that students relied on contextual clues to find the necessary portions of the textbook or other course materials relevant to the question rather than their understanding of the assigned readings. Students may have scored more highly on human-authored MCQs because they were attuned to the instructor's approach to course content and style of questioning.

\section{Limitations}

LLM-based tools differ in implementation, parameters, and training data. It is plausible that results of this study are dependent in some part on the specific LLM-based tool used to generate the questions and the training data used to construct the model used for similarity comparisons. Our choice of ChatGPT-4o for constructing MCQs is informed by work that suggests our choice of a commercial LLM should outperform open source models~\cite{ateia2024can}. 

Cosine similarities were evaluated using only SBERT. We did not consider course materials other than the required textbook when computing cosine similarities. Course materials included other suggested books, online articles, code samples, lecture slides, as well as recordings of the synchronous online lectures. We also did not fine tune the LLM to the textbook itself, but rather relied on the general model for distance calculations. Lastly, regardless of an MCQ's authorship, prior work suggests that an inherent limitation of MCQs is their inability to assess activities in the "create" (or "synthesis") domain of Bloom's taxonomy~\cite{Krathwohl01112002}.

\section{Future Work}

In our experience, most LLM-authored MCQs failed validation and were not used. We ensured that the questions that were ultimately given to students were stated clearly, with unambiguous directions and assumptions, plausible choices, and exactly one correct choice. We also ensured that the expected effort to arrive at the correct answer was similar to the expectation implicit in the instructor's own question. This last criterion is perhaps the most subjective one and another instructor may approach a solution differently. Replicating this study with another instructor and another course would provide the necessary evidence to determine the effect of the instructor's judgement about the quality of the LLM-authored MCQs on the scores attained on those same questions.

In general, this and all similarly constructed studies should be reproduced to control for the influence of the instructor's judgement used in constructing and validating questions. Likewise, a study such as this should be conducted in a different context - any other discipline, any other course, any other instructor. Some qualities of LLM-generated text within a knowledge domain could be representative of the availability of training data. Just as English-language text is far more prevalent online, and so proportionally better represented in LLMs, this also may be the case for Computer Science-related text. Replicating the study in other disciplines could reasonably yield different results. 

%% file: conclusion.tex
\section{Conclusion}

Based on assessment data collected from 25 participants throughout an 8-week undergraduate summer course on operating systems, we find that assessment questions written with the aid of an LLM, like ChatGPT, have no superficial qualities that distinguish them from solely human-authored questions. However, we find that students score significantly lower on assessment questions authored with the aid of an LLM. We also find a significant correlation between cosine similarity of the human questions and textbook compared to the cosine similarity of the LLM-aided questions and textbook -- the LLM produced questions that were on average closer to the textbook according to cosine similarity. 
This result may suggest that despite the fact that the LLM-aided questions were well informed and had the correct information at a similar level to the instructor, the instructor's teaching and writing style may implicitly influence student performance.
%It is important to note that while LLM-based tools like ChatGPT have the potential to free instructors from some of the drudgery of constructing novel multiple-choice exam questions for well-worn material, their initial output is seldom immediately usable and sometimes not at all. All output must be validated by a human expert to ensure that the question is well-formed, that the correct answer is indeed present among the given choices, that the given choices are plausible and do not represent hints, and that the expected skills, effort, and knowledge are adequately represented when deriving a solution. 